\documentclass[
    ,final            
  ]
  {aipproc}

\layoutstyle{8x11double}

\usepackage{amsmath}
\usepackage{braket}

\newcommand{\p}{\partial}
\renewcommand{\d}{\ensuremath{\mathrm{d}}}
\newcommand{\ii}{\ensuremath{\mathrm{i}}}

\newcommand{\FP}{\ensuremath{\mathrm{FP}}}

\newcommand{\GZ}{\ensuremath{\mathrm{GZ}}}

\begin{document}

\title{From propagators to glueballs in the Gribov-Zwanziger framework}

\classification{11.10.Gh, 11.15.Tk}
\keywords      {Gribov-Zwanziger, renormalization, propagators, bound states}

\author{Nele Vandersickel}{
  address={Ghent University, Department of Physics and Astronomy \\
                            Krijgslaan 281-S9, B-9000 Gent, Belgium}
}

\author{David Dudal}{
  address={Ghent University, Department of Physics and Astronomy \\
                            Krijgslaan 281-S9, B-9000 Gent, Belgium}
}

\author{Orlando Oliveira}{
  address={Departamento de F\'{\i}sica, Universidade de Coimbra,\\
   P-3004-516 Coimbra, Portugal}
}

\author{Silvio P. Sorella}{
  address={ Departamento de F\'{\i }sica Te\'{o}rica, Instituto de F\'{\i }sica, UERJ - Universidade do Estado do Rio de Janeiro\\
                            Rua S\~{a}o Francisco Xavier 524, 20550-013 Maracan\~{a}, Rio de Janeiro, Brasil}
}

\begin{abstract}
Over the last years, lattice calculations in pure Yang-Mills gauge theory seem to have come more or less to a consensus. The ghost propagator is not enhanced and the gluon propagator is positivity violating, infrared suppressed and non-vanishing at zero momentum. From an analytical point of view, several groups are agreeing with these results. Among them, the refined Gribov-Zwanziger (RGZ) framework also accommodates for these results. The question which rises next is, if our models hold the right form for the propagators, how to extract information on the  real physical observables, i.e.~the glueballs? How do the operators which represent glueballs look like? We review the current status of this matter within the RGZ framework.
\end{abstract}

\maketitle


\section{Introduction on the GZ formalism}
As is well known, the Yang-Mills action needs to be gauge fixed in order to define the path integral. One way to do this is trough the gauge fixing proposed by Faddeev and Popov. However, as is shown by Gribov, this gauge fixing suffers from Gribov copies \cite{Gribov:1977wm}. If we take e.g.~the Landau gauge, $\partial_\mu A_\mu = 0$, it is very easy to show that there exist gauge equivalent fields $A'_\mu$ also fulfilling the Landau gauge $\p_\mu A_\mu'$ when having zero modes of the Faddeev-Popov operator
\begin{eqnarray*}
 \mathcal M^{ab} &=&  - \p_\mu D_\mu = -\p_\mu (\p_\mu \delta^{ab} - g f^{abc} A_\mu^c)\;.
\end{eqnarray*}
Therefore, Gribov proposed to restrict the region of integration to $\Omega$
\begin{eqnarray*}
\Omega=\left\{  A_{\mu}|\partial_{\mu} A_{\mu}=0, \mathcal M>0  )    \right\}\;,
\end{eqnarray*}
the region which is free of zero modes of the Faddeev-Popov operator $\mathcal M^{ab}$. We should mention however, that there are still Gribov copies inside the Gribov region \cite{vanBaal:1991zw}. In 1989, Zwanziger implemented this restriction to the Gribov region to all orders: the Gribov-Zwanziger action was born \cite{Zwanziger:1989mf}. Let us here immediately present the localized version of this action, namely
\begin{equation*}
S_\GZ = S_{\FP} + S_0+ S_{\gamma}\;,
\end{equation*}
with $S_{\FP}$ the usual Faddeev-Popov action,
\begin{equation}
S_{\FP} = \frac{1}{4}\int \d^4 F_{\mu\nu}^a F_{\mu\nu}^a +  \int \d^4 x \left( b^{a}\partial_\mu A_\mu^{a}+\overline{c}^{a}\partial _{\mu } D_{\mu}^{ab}c^b \right) \;,
\end{equation}
and $S_0$ and $S_\gamma$ given by
\begin{align}
S_0 &= \int \d^{4}x\left( \overline{\varphi}_\mu^{ac} \mathcal M^{ab}\varphi^{bc}_\mu -\overline{\omega}_{\mu}^{ac} \mathcal M^{ab} \omega_\mu^{bc} \right) \;, \nonumber\\
S_{\gamma}&= -\gamma ^{2}g\int\d^{4}x\left( f^{abc}(\varphi _{\mu }^{bc} +\overline{\varphi }_{\mu }^{bc} )A_{\mu }^{a} + \frac{d}{g}\left(N^{2}-1\right) \gamma^{2} \right) \;. \nonumber
\end{align}
The fields $( \overline{\varphi }_{\mu}^{ac},\varphi_{\mu}^{ac}) $ are a pair of complex conjugate bosonic fields, while $( \overline{\omega }_{\mu}^{ac},\omega_{\mu}^{ac}) $ are anticommuting fields. $\gamma$ is not free, but fixed by a horizon condition, $\Braket{ gf_{cka}A_{\mu}^{k}( \varphi_{\mu}^{ac}+\bar{\varphi}_{\mu}^{ac} ) } = 2\gamma^{2}4(N^{2}-1)$. Looking at this horizon condition, we see that it is equivalent with a dimension two condensate. Therefore, we could investigate other dimension two condensates, namely $\braket{\overline\varphi\varphi-\overline\omega\omega}$ and $\braket{A^2}$. Including these condensates gives rise to the refined GZ (RGZ) action \cite{Dudal:2008sp}. We have shown that this can be done in a renormalizable way. We stress that the masses related to the condensates respectively $M^2$ and $m^2$ are dynamically generated.

\section{Propagators}
Let us now look at the ghost and the gluon propagator. Firstly, it is sure that perturbation theory fails in the infrared region as perturbatively, the gluon propagator blows up, while all lattice calculations show that there is infrared suppression. Secondly, the Gribov-Zwanziger predicts a gluon propagator which is infrared suppressed, and therefore clearly goes beyond perturbation theory. Unfortunately, at zero momentum, this gluon propagator vanishes, which contradicts the current numerical simulations that this propagator is non-vanishing at zero momentum \cite{Cucchieri:2007md,Bogolubsky:2007ud}. In addition, the ghost propagator shows infrared enhancement, which is also no longer observed in the infrared. Therefore, something is still missing in the GZ framework. Thirdly, let us investigate the refined GZ action. In this case, the tree level gluon propagator behaves like
\begin{equation}\label{nice}
D(p^2) = \frac{p^2 + M^2}{p^4 + \left(M^2 + m^2\right) p^2 + 2 g^2 N \gamma^4 + M^2 m^2  }\;.
\end{equation}
This propagator is clearly infrared suppressed, also violating positivity and is no longer zero at zero momentum. Also the ghost propagator is no longer enhanced as an effect of the condensates. This agrees again qualitatively with the latest lattice results. \\
In \cite{Dudal:2010tf}, we have investigated our specific form of the gluon propagator on the lattice. Two main conclusions could be drawn: 1) If $m^2 =0$, the fits are of low quality, therefore, the condensate $\braket{A^2}\sim m^2$ is indispensable; 2) Nice fits for the form \eqref{nice} were found when including the condensate $\Braket{A^2}$. Let us recall some estimates: $M^2 = 2.14 \pm 0.13 $GeV$^2$, $
m^2 = -1.78 \pm 0.14$ GeV$^2$ and $D(0) = 8.2 \pm 0.5$ GeV$^2$. From the estimate of $m^2$, we find that $\braket{g^2A^2}_{\mu=10\text{GeV}}\approx 3 \text{GeV}^2$ which is in the same ballpark as other, rather independent, approaches \cite{Boucaud:2008gn,RuizArriola:2004en}.\\
In conclusion, we could state that the RGZ framework provides a possible explanation of the behavior of the gluon and the ghost propagator. In fact, one can show that more condensates can be present, and alternative RGZ frameworks exist \cite{nele,Gracey:2010cg}. \\
Let us also stress that in 2d, the ghost propagator is still enhanced, while the gluon propagator is zero at zero momentum, \cite{Cucchieri:2007rg}. We have also provided evidence that in 2d, the GZ formalism still holds the right results as refinement is impossible in 2d \cite{Dudal:2008xd}.

\section{The quest for physical operators}
Now that the propagators match with the lattice, we can wonder whether the RGZ also holds information about the particles of (quenched) QCD? The idea to proceed is the following: we want to find an operator $\mathcal O$, so that the correlator $\Braket{\mathcal O(k) \mathcal O(-k)}$ can be put into a spectral representation:
\begin{align*}
 \Braket{ \mathcal O(k) \mathcal O(-k) }  =  \int_{\tau_{0}}^{\infty} \d\tau \; \rho({\tau}) \; \frac{1}{\tau+k^2} \;.
\end{align*}
We then introduce $ F(z) =    \int_{\tau_{0}}^{\infty} \d\tau \; \rho(\tau) \; \frac{1}{\tau+z}$ so that when going to Minkowski space, i.e. $k^2_{Eucl} \rightarrow - k^2_{Mink}$, we find a discontinuity along the positive real axis. In addition we want $\rho({\tau})$ to be positive for $\tau \geq \tau_0$ in order to give a particle interpretation to the correlator\footnote{$\rho$ is proportional to the cross section, and thus has to be positive.} and  obviously $\mathcal O(k)$ has to be renormalizable.\\
We have worked out two different angles to attack this problem. In the first approach, we have introduced $i$-particles \cite{Baulieu:2009ha}
\begin{eqnarray*}
  \lambda_\mu^a&=&\frac{1}{\sqrt{2}}A_\mu^a+\frac{\ii}{2\sqrt{N}}f^{abc}\left(\varphi_\mu^{bc}+\overline \varphi_\mu^{bc}\right)\;,   \nonumber\\
\eta_\mu^a&=&\frac{1}{\sqrt{2}}A_\mu^a-\frac{\ii}{2\sqrt{N}}f^{abc}\left(\varphi_\mu^{bc}+\overline \varphi_\mu^{bc}\right)\;,
\end{eqnarray*}
to make the quadratic part of the action diagonal. With the $i$-field strengths, $\lambda^{a}_{\mu\nu} = \partial_{\mu}  \lambda^{a}_{\nu} -  \partial_{\nu}  \lambda^{a}_{\mu} $ and $\eta^{a}_{\mu\nu}  = \partial_{\mu}  \eta^{a}_{\nu} -  \partial_{\nu}  \eta^{a}_{\mu}$, we have proposed the following operator
\begin{align*}
O^{(1)} &=  \left( \lambda^{a}_{\mu\nu}(x) \eta^{a}_{\mu\nu}(x) \right) \;.
\end{align*}
The corresponding correlator $\braket{O^{(1)}(k)O^{(1)}(-k)}$ can be put in a spectral representation:
\begin{equation*}
\braket{O^{(1)}(k)O^{(1)}(-k)} = \int_{2\lambda^2}^{\infty} \d \tau \rho(\tau) \frac{1}{\tau + k^2}\;,
\end{equation*}
whereby $ \rho(\tau) \geq 0$. Unfortunately, the operator $O^{(1)}$ is not renormalizable.\\
Let us therefore also discuss the second angle. In \cite{Dudal:2009zh}, we have investigated the renormalization of the typical scalar glueball operator $F^2$ with the GZ action. Let us start by noting that the Faddeev-Popov action $S_\FP$ is invariant under the BRST transformation $s$: $s S_\FP =0$,
\begin{align*}
sA_{\mu }^{a} &=-\left( D_{\mu }c\right) ^{a}\;, & sc^{a} &=\frac{1}{2}gf^{abc}c^{b}c^{c}\;, \\
 s\overline{c}^{a} &=b^{a}\;,&   sb^{a}&=0\;.
\end{align*}
This BRST invariance lies at the origin of the Slavnov-Taylor identity, which allows us to prove the renormalizability. Moreover, the BRST charge allows us to define the sub-space of the physical states and to establish the unitarity of the $S$ matrix. Unfortunately, the GZ action is no longer invariant under the BRST transformation $s$. The new fields transform as
\begin{align}
s\varphi _{i}^{a} &=\omega _{i}^{a}\;,&s\omega _{i}^{a}&=0\;, & s\overline{\omega }_{i}^{a} &=\overline{\varphi }_{i}^{a}\;,& s \overline{\varphi }_{i}^{a}&=0 \;,
\end{align}
and thus the breaking is proportional to $\gamma$, $sS_\GZ = s \left(S_{\mathrm{YM}}+S_{gf} + S_0+ S_{\gamma}\right)  = s \left(S_\gamma \right)  \sim  \gamma^2 \not= 0$. Despite this breaking, the GZ action is still renormalizable, due to a rich set of Ward identities. Moreover, only two renormalization constants are needed, which is the same as in the Yang-Mills theory. Let us also mention that by introducing extra fields, a symmetry can be restored again \cite{Dudal:2010hj}. With the breaking in mind, we can renormalize $F^2$ within the GZ framework as done in \cite{Dudal:2009zh}. This renormalization is however far from trivial due to the breaking of the BRST symmetry. We find that $F^2$ mixes with the following operator:
\begin{equation}
\mathcal E = s( \ldots) + \gamma ^{2}  D_\mu^{ab} \left( \varphi_\mu^{ba} + \overline \varphi_\mu^{ba} \right) + d (N^2 - 1) \gamma \;.
\end{equation}
Moreover, we have constructed a renormalization group invariant given by
\begin{equation}
O^{(2)} =   \frac{\beta(g^2)}{g^2} F^2 - 2 \gamma_c \mathcal E \;.
\end{equation}
Therefore, we would propose the following correlator
\begin{align}
&\Braket{O^{(2)}(x) O^{(2)}(y)} = \nonumber\\
 &\Braket{   \left[ \frac{ \beta} {g^2} F^2  -2 \gamma_c \mathcal E \right](x)   \left[ \frac{ \beta} {g^2} F^2  -2 \gamma_c \mathcal E \right](y)} \nonumber\\
                                 &\not= \left( \frac{ \beta} {g^2}\right)^2 \Braket{  F^2(x) F^2(y) }\;.
\end{align}
Notice that the breaking of the BRST symmetry is the reason that this correlator contains extra terms beside $\Braket{  F^2(x) F^2(y) }$. In ordinary Yang-Mills theory, a similar correlator can be found, but it reduces the $\braket{F^2(x) F^2(y)}$ due to the presence of the BRST symmetry. Unfortunately, with this correlator, we are unable to find a good spectral representation. This was already found in \cite{Zwanziger:1989mf}:
\begin{equation*}
 \int d^{4}x\ e^{-ikx\ }\Braket{ F^{2}(x)F^{2}(0)} = G^{\rm phys}(k^2) + G^{\rm unphys}(k^2) \; ,
\end{equation*}
whereby the unphysical part, $G^{\rm unphys}(k^2)$, displays cuts along the imaginary axes beginning at the unphysical values $k^2=\pm 4i{ \gamma}^2$ and the physical part, $G^{\rm phys}(k^2)$, has a cut beginning at the physical threshold $k^2=-2{\gamma}^2$. \\
\\
For further clarification, let us show the relation between $\mathcal O^{(1)}$ and $\mathcal O^{(2)}$. At lowest order, we have that
\begin{equation*}
\underbrace{F^2_{\mu\nu}}_{O^{(2)}} =  \underbrace{\lambda_{\mu\nu}^a\eta_{\mu\nu}^a}_{O^{(1)}} + 1/2\lambda_{\mu\nu}^a\lambda_{\mu\nu}^a+1/2\eta_{\mu\nu}^a\eta_{\mu\nu}^a \;,
\end{equation*}
and thus it are in fact the two last terms which cause the bad spectral representation in $\braket{F^2(x) F^2(y)}$, and the question remains how we can find a renormalizable operator with a good spectral representation.\\
We can already mention that so far, we have only investigated spectral representations within the GZ framework. The hope is that within the RGZ framework, we can still find an operator with the required properties as some aspects change of the analysis. Also, the study of glueballs looks very promising within the (R)GZ framework, see, for example, the recent results in \cite{Dudal:2010cd}. This question of finding descent operators is not only relevant within the Gribov-Zwanziger context, but for all people involved in infrared propagators QCD. How to find good spectral operators starting from unphysical operators?


\begin{theacknowledgments}
  D.~Dudal and N.~Vandersickel are  supported by the Research Foundation-Flanders (FWO). The work of O.~Oliveira is supported by FCT under project CERN/FP/83644/2008. Silvio P. Sorella acknowledges support from CNPq-Brazil, Faperj, SR2-UERJ, CAPES and CLAF.
\end{theacknowledgments}



\bibliographystyle{aipproc}   

\begin{thebibliography}{9}
\bibitem{Gribov:1977wm}
  V.~N.~Gribov,
  Nucl.\ Phys.\ B {\bf 139} (1978) 1.

\bibitem{vanBaal:1991zw}
  P.~van Baal,
  Nucl.\ Phys.\  B {\bf 369} (1992) 259.

\bibitem{Zwanziger:1989mf}
  D.~Zwanziger,
  Nucl.\ Phys.\  B {\bf 323} (1989) 513.


\bibitem{Dudal:2008sp}
  D.~Dudal, J.~A.~Gracey, S.~P.~Sorella, N.~Vandersickel and H.~Verschelde,
  Phys.\ Rev.\  D {\bf 78} (2008) 065047.

\bibitem{Cucchieri:2007md}
  A.~Cucchieri and T.~Mendes,
  PoS {\bf LAT2007} (2007) 297
  [arXiv:0710.0412 [hep-lat]].

\bibitem{Bogolubsky:2007ud}
  I.~L.~Bogolubsky, E.~M.~Ilgenfritz, M.~Muller-Preussker and A.~Sternbeck,
  PoS {\bf LAT2007}, 290 (2007)
  [arXiv:0710.1968 [hep-lat]].

\bibitem{Dudal:2010tf}
  D.~Dudal, O.~Oliveira and N.~Vandersickel,
  Phys.\ Rev.\  D {\bf 81} (2010) 074505
  [arXiv:1002.2374 [hep-lat]].


\bibitem{Boucaud:2008gn}
Ph.~Boucaud, F.~De Soto, J.~P.~Leroy, A.~Le Yaouanc, J.~Micheli, O.~Pene and J.~Rodriguez-Quintero, Phys.\ Rev.\  D {\bf 79} (2009) 014508.


\bibitem{RuizArriola:2004en}
E.~Ruiz Arriola, P.~O.~Bowman and W.~Broniowski, Phys.\ Rev.\  D {\bf 70} (2004) 097505.

\bibitem{nele}
work in progress

\bibitem{Gracey:2010cg}
  J.~A.~Gracey,
  arXiv:1009.3889 [hep-th].

\bibitem{Cucchieri:2007rg}
  A.~Cucchieri and T.~Mendes,
  Phys.\ Rev.\ Lett.\  {\bf 100}, 241601 (2008).

\bibitem{Dudal:2008xd}
  D.~Dudal, S.~P.~Sorella, N.~Vandersickel and H.~Verschelde,
  Phys.\ Lett.\  B {\bf 680}, 377 (2009).

\bibitem{Baulieu:2009ha}
  L.~Baulieu, D.~Dudal, M.~S.~Guimaraes, M.~Q.~Huber, S.~P.~Sorella, N.~Vandersickel and D.~Zwanziger,
  Phys.\ Rev.\  D {\bf 82} (2010) 025021.

\bibitem{Dudal:2009zh}
  D.~Dudal, S.~P.~Sorella, N.~Vandersickel and H.~Verschelde,
  JHEP {\bf 0908} (2009) 110.

\bibitem{Dudal:2010hj}
  D.~Dudal and N.~Vandersickel,
  arXiv:1010.3927 [hep-th].

\bibitem{Dudal:2010cd}
  D.~Dudal, M.~S.~Guimaraes and S.~P.~Sorella,
  arXiv:1010.3638 [hep-th].


\end{thebibliography}

\end{document}